\documentclass[letterpaper]{article} 
\usepackage{aaai24}  

\usepackage{times}  
\usepackage{helvet}  
\usepackage{courier}  
\usepackage[hyphens]{url}  
\usepackage{graphicx} 
\usepackage{natbib}  
\usepackage{caption} 
\usepackage{algorithm}
\usepackage{algorithmic}

\usepackage{multirow} 
\usepackage[table]{xcolor} 
\usepackage{enumitem} 
\usepackage{booktabs} 
\usepackage{tabularx} 
\usepackage{array}    
\usepackage{tikz}
\usetikzlibrary{shapes, arrows.meta, positioning}
\usepackage{xcolor}
\usepackage{tabularx}
\usepackage{newfloat}
\usepackage{listings}
\DeclareCaptionStyle{ruled}{labelfont=normalfont,labelsep=colon,strut=off} 
\floatstyle{ruled}
\newfloat{listing}{tb}{lst}{}
\floatname{listing}{Listing}
\title{Human-Centered AI Applications for \\Canada's Immigration Settlement Sector}

\author {
    Isar Nejadgholi\textsuperscript{\rm 1},
    Maryam Molamohammadi\textsuperscript{\rm 2},
    Kimiya Missaghi\textsuperscript{\rm 3,\rm 4}
    Samir Bakhtawar\textsuperscript{\rm 5}
}
\affiliations {
    \textsuperscript{\rm 1}National Research Council Canada\\
    \textsuperscript{\rm 2}MILA, Quebec Institute of Artificial Intelligence\\
    \textsuperscript{\rm 3}University of Ottawa\\
    \textsuperscript{\rm 4}PeaceGeeks\\
    \textsuperscript{\rm 5}Independent Researcher\\
}

\usepackage{bibentry}

\begin{document}

\maketitle

\begin{abstract}
While AI has been frequently applied in the context of immigration, most of these applications focus on selection and screening, which primarily serve to empower states and authorities, raising concerns due to their understudied reliability and high impact on immigrants' lives. In contrast, this paper emphasizes the potential of AI in Canada’s immigration settlement phase, a stage where access to information is crucial and service providers are overburdened. By highlighting the settlement sector as a prime candidate for reliable AI applications, we demonstrate its unique capacity to empower immigrants directly, yet it remains under-explored in AI research. We outline a vision for human-centred and responsible AI solutions that facilitate the integration of newcomers. We call on AI researchers to build upon our work and engage in multidisciplinary research and active collaboration with service providers and government organizations to develop tailored AI tools that are empowering, inclusive and safe.

\end{abstract}

\section{Introduction} 
\label{sec:intro}

Canada is recognized as a multicultural country known for its inclusion-driven approach, actively inviting people from diverse backgrounds to contribute to its economic, social, and cultural development. According to a new Statistics Canada report, in 2023, Canada was one of the fastest-growing countries in the world, with immigration being the main driver of population growth \cite{government_of_canada_daily_2023}. Every year, the government of Canada sets targets for each immigration category, including economic programs, family reunification, and humanitarian programs. Additionally, a large population of temporary residents migrate to Canada every year to work or study. Canada's immigration process is implemented in three main steps: 1) \textit{Selection}, which aims to distribute the benefits of immigration across the country; 2) \textit{Screening}, which ensures newcomers do not pose a risk to Canada, are in good health, and haven't committed a serious crime or violated any laws or human rights; 3) \textit{Settlement\footnote{ ``Settlement'' is an umbrella term. ``Resettlement'' is the precise term for the settling process of refugees and forced immigrants.  } and Integration}, which provides services to help newcomers adapt to life in Canada and contribute to its prosperity \cite{Immigrat26:online}. 


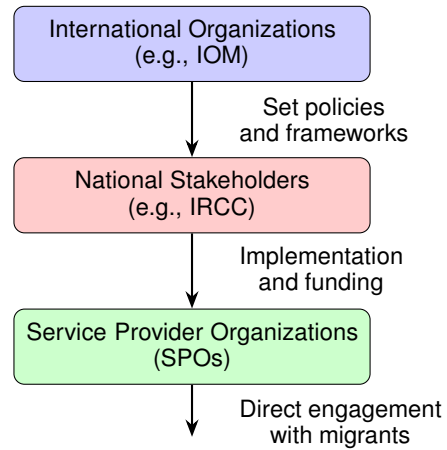
\begin{figure}[ht]
\centering

\begin{tikzpicture}[node distance=1cm and 1.2cm, 
                    auto, 
                    every node/.style={align=center, font=\sffamily\small}, 
                    arrow/.style={-{Stealth[]}, thick},
                    box/.style={rectangle, draw, rounded corners, fill=blue!20, minimum height=1cm, text width=4.5cm, align=center, font=\sffamily\small}]
  \node (international) [box, fill=blue!20] {International Organizations\\(e.g., IOM)};
  \node (national) [box, fill=red!20, below=of international] {National Stakeholders\\(e.g., IRCC)};
  \node (sporganizations) [box, fill=green!20, below=of national] {Service Provider Organizations\\(SPOs)};

  \draw[arrow] (international) -- (national) node[midway, fill=white, xshift=0.5cm] {Set policies\\and frameworks};  
  \draw[arrow] (national) -- (sporganizations) node[midway, fill=white, xshift=0.5cm] {Implementation\\and funding};  
  \draw[arrow] (sporganizations) -- +(0, -1.2) node[near end, right=0.5cm] {Direct engagement\\with migrants};  

\end{tikzpicture}
\caption{Immigration Governance Structure}
\label{fig:immigration_governance}
\end{figure}

As shown in Figure \ref{fig:immigration_governance}, immigration is governed by multiple levels of international and national actors \cite{bakhtawar2022artificial}. International organizations, such as the International Organization for Migration (IOM), set broad policies and frameworks, facilitating cooperation across borders. Further, national stakeholders, such as Immigration, Refugees, and Citizenship Canada (IRCC), craft and implement national policies, provide funding, and oversee the integration of migrants into the country's social fabric. Moreover, community-based and government-funded Service Provider Organizations (SPOs) are on the frontline, directly engaging with migrants to provide personalized support and services. 

In this work, we focus on the potential implications of AI technologies in the \textbf{settlement} sector. The Canadian government recognizes that in order to realize the economic, social, and cultural benefits of immigration, newcomers must integrate successfully and in a timely manner into Canadian society. Hence, government-sponsored settlement programs assist immigrants and refugees in tackling the challenges unique to them \cite{immigration_settlement_2022}. Settlement services are delivered by SPOs
 and help newcomers learn about life in Canada, improve their language skills, enter the job market, and make connections, all of which require timely access to accurate and complete information \cite{allard2022so}. Despite significant investments, the Canadian settlement sector faces operational bottlenecks due to the government’s increasing immigration targets \cite{immigration_notice_2023}. With growing demand, the Canadian settlement sector is evidently stretched and needs enhanced efficiency. 
 Settlement services, by nature, are information-heavy, requiring extensive data processing to evaluate the assets and needs of newcomers and refer them to the appropriate resources. Therefore, reliable AI solutions tailored to the needs of newcomers might ease the pressure on the sector. Furthermore, these solutions can be integrated into online platforms, which are preferred by clients seeking \textit{Information and Orientation (I\&O)} services and have shown stronger knowledge outcomes \cite{part2dig35:online}.



Motivated by the significance of successful newcomer integration and the increasing demand for efficient information processing in the settlement sector, we aim to explore the potential and limitations of AI solutions in this space. For that, we engaged with a broad spectrum of stakeholders in the sector through interviews, participated in national conferences and events, and identified categories of needs that could be alleviated through the integration of reliable AI tools. We then scanned research works that applied AI tools to immigration. Notably, we observed a significant gap in AI research on settlement processes. To address this gap, we mapped the tasks performed by this sector to well-established AI tasks and applications. We also provide recommendations for the responsible integration of AI tools into the settlement sector. Specifically, we underscore a significant gap in multidisciplinary studies that merge AI capabilities with other domain-specific expertise to develop reliable and customized solutions for this critical sector. We call on AI researchers to develop responsible AI tools that can be integrated into pre-existing service structures in a safe and inclusive manner. 

\vspace{5pt}
Our main contributions are as follows: 

\begin{itemize}
\item Through engaging with multiple stakeholders, including community organizers, SPOs, and academics, we identified the settlement sector as a prime candidate for responsible AI applications. 
\item Through reviewing the AI literature, we identified gaps in AI research related to the settlement sector. 
\item Guided by the service types provided by Canadian SPOs, we mapped the needs of the settlement sector to AI applications and core tasks.
\item Finally, we provided recommendations on how to foster multidisciplinary AI research to alleviate the challenges faced by the settlement sector. 
\end{itemize}

\section{Community Engagements and its Findings}

We took a multidisciplinary and participatory approach and aimed to understand the settlement space and its complexities and evaluate the feasibility of introducing AI-enabled tools responsibly. Our team is comprised of researchers from various backgrounds: an applied AI researcher, a responsible AI advisor and researcher, and two researchers who extensively and regularly work with the settlement sector in both academic and administrative capacities. 

We started our search by conducting ten flexible and open-ended interviews and consultations with the stakeholders, including community organizers, SPO members, IRCC employees, immigrant-serving communication consultants, and academics in this field. Our main goal in these interviews was to understand the current state of the settlement sector, its existing workflows and organizational structures, which are reflected in this paper. We then participated in and held a workshop at the 2023 Pathway to Prosperity National Conference \cite{P2P}. The Pathways to Prosperity (P2P) Partnership 
includes key federal and provincial/territorial migration ministries; municipalities; national, regional, and local organizations involved in newcomer settlement; and researchers from across Canada. The main activities of the Partnership are primary and secondary research, knowledge transfer, education, and mutual learning. Recognizing that forming a multidisciplinary team and enlisting domain experts are essential parts of responsible AI and meaningful adaptation, our objectives for this workshop were to: 1) Evaluate our findings from literature reviews and interviews by engaging with a wider group within this community, 2) Broaden our understanding of the existing needs, 3) Assess the potential integration of AI-enabled tools and 4) Find collaborator(s) already active in this field. After interviews, the workshop, and the onboarding of two members from the public policy and administration domain who specialized in settlement services, we continued the literature review and brainstorming ideas, proposing a mapping of needs to AI applications with a multidisciplinary lens. 

This process revealed a significant operational bottleneck in the sector, specifically in information-intense tasks. Despite the sector's strong interest and willingness to embrace AI to enhance performance and efficiency, we observed a significant gap in educational resources to support AI literacy within this sector. We noted instances of impressive and successful integration of AI in SPO operations; however, these applications were limited in scope and would benefit from a more systematic adoption framework. We also observed an organic and ongoing research collaboration between the settlement sector and social scientists, although collaborations with AI researchers are missing.

Moreover, there is frequently a lack of digital infrastructure for constructing meaningful databases and concrete guidelines for curating, structuring, and assessing available data. There appears to be minimal transversal collaboration and knowledge transfer across different levels of government to support such data management infrastructures. Among immigrants and SPOs, there's a growing trend towards ad-hoc use and reliance on ChatGPT-like systems, raising questions about the reliability of these tools for specific recommendations \cite{FAccT}.
Learning more about the SPOs' day-to-day operations, we concluded that many tasks within this sector can align with existing ML-based tools, facilitating accurate and relevant information delivery, resource allocation, and referrals. By embracing a human-centred design ethos and incorporating a fair degree of human-in-the-loop interaction, desirable outcomes can be achieved.

\section{Current Landscape of AI Research in Immigration}

\begin{table*}[ht]
\centering
\small
\begin{tabularx}{\textwidth}{@{}p{2cm} p{5cm} p{4.6cm} p{4.6cm}@{}}
\toprule
\textbf{} & \textbf{Data Analytics} & \multicolumn{2}{c}{\textbf{Customized AI Tools}} \\ 
\midrule
\textbf{International Organizations} &
\begin{itemize}[leftmargin=*]
  \item Analysis of global migration trends 
  \item Analysis of stereotypes in the visual portrayal of immigrants
  \item Anti-immigrant hate speech analysis
  \item Impact of COVID-19 on migration and mobility
  \item Public attitudes measurement
  \item Immigration policy analysis
\end{itemize} &
\begin{itemize}[leftmargin=*]
  \item Machine translation  
  \item Migration and forced migration prediction
  \item Refugee crises prediction 
  \item Anti-immigration speech detection 
  \item Refugee movement prediction
\end{itemize} &
\begin{itemize}[leftmargin=*]
  \item Migration management automation
  \item Forecasting integration trends
  \item Population growth prediction
  \item Humanitarian needs assessments 
  \item Migrant smuggling prevention
  \item Legal assistant for refugees
\end{itemize} \\ 
\midrule
\textbf{National \newline Government} &
\begin{itemize}[leftmargin=*]
  \item National migration data analysis
  \item Social media mining of migration discourse
  \item Studying mental health among immigrants
  \item Studying the migration and asylum policies
  \item Asylum claim analysis
  \item Anti-immigrant hate speech analysis
  \item Political speech analysis
\end{itemize} &
\begin{itemize}[leftmargin=*]
  \item Machine translation for forced migration 
  \item Migration modeling and forecasting 
  \item Border security management
  \item Asylum adjudication support 
  \item Communication of migration policies
  \item Illegal immigration prediction 
  \item Medical diagnosis
  \item Migrants' income prediction
  \item Fake news detection for migrants
\end{itemize} &
\begin{itemize}[leftmargin=*]
  \item Determining occupational classification codes
  \item Identify and track refugees in surveillance zones for humanitarian aid
  \item AI for legal text processing 
  \item Migrants' population prediction 
  \item Refugee and asylum claim processing 
  \item Social media content summarization for inclusive policy-making
  \item Automated asylum seeker interviews
\end{itemize} \\ 
\midrule
\textbf{Settlement and \newline Resettlement} &
\begin{itemize}[leftmargin=*]
  \item Social media content analysis  
  \item Migrant children's education
  \item Immigrant integration analysis
\end{itemize} &
\begin{itemize}[leftmargin=*]
  \item Migrant skill matching
  \item Public services accessibility 
  \item Identification of vocational domains
  \item Health literacy estimation
\end{itemize} &
\begin{itemize}[leftmargin=*]
  \item Integration support 
  \item Personal assistant for migrants
  \item Overcoming language barriers 
  \item Employment-related skill training 
\end{itemize} \\ 
\bottomrule
\end{tabularx}
\caption{Examples of AI research in immigration-related applications. While many research works have applied AI tools and techniques to assist international and national stakeholders, the potential of AI in settlement processes is underexplored.}
\label{tab:ai_literature}
\end{table*}

Here, we explore the current scope of AI research applied to challenges associated with immigration. Our analysis is structured along two dimensions that categorize AI applications based on their functional roles and applicability across different governance levels. Along the first dimension, we classify AI solutions into three distinct categories to reflect the specific operational capacities in which AI technologies are employed: 1) AI-based data analytics for drawing insights from large datasets, 2) Customized AI tools specifically tailored to assist 
in decision-making processes, 
and 3) General-purpose AI-enabled tools available to the public. Further, we consider a second dimension, which categorizes the literature based on three levels of governance, where these applications can be implemented by 1) International stakeholders, 2) Federal organizations, or 3) Settlement service providers. 

\vspace{5pt}
\noindent \textbf{Selection of Research Papers}: For our search, we used the Google Scholar API  with a set of immigration-related keywords -- \textit{newcomer, immigration, migration, asylum seeker, immigrant, migrant, refugee} -- and a set of AI-related keywords -- \textit{neural network, deep learning, machine learning, computer vision, natural language processing, AI, artificial intelligence} -- and retrieved papers which included at least one of the terms from immigration-related keywords and at least one of the AI-related terms. We limited our search to the time period of January 2020 to April 2024. This search retrieved 1362 papers, for which we collected the title and a snippet that included searched keywords. We then used the GPT4 API \cite{achiam2023gpt} to identify if these papers are, in fact, applying AI models to immigration-related problems. After filtering non-relevant papers, we were left with 242 papers which apply AI to address an immigration-related problem. We manually checked these papers and categorized them across the dimensions of functional role and level of governance described above. We also complemented this search with a manual search of related papers. The list of applications we found in this search is presented in Table \ref{tab:ai_literature} and discussed in the following. General-purpose AI tools are not pictured in this table due to a lack of relevant research but will be elaborated on later.   

\vspace{5pt}
\noindent \textbf{AI for Data Analytics and Insight Generation:} This category includes AI algorithms and tools used to extract insights from large datasets, and uncover patterns and relationships that might not be sufficiently understood through manual data inspection or statistical analysis. As shown in Table \ref{tab:ai_literature}, the research works we found in this category typically focus on social network analysis for opinion mining, trend analysis in migration patterns, economic impact studies, and population-level healthcare needs assessments. The generated knowledge contributes to strategic policy formation and long-term planning.

A large body of AI research that applies data analytics techniques to migration data has potential implications at the international level of governance \cite{beduschi2021international}. For example, \citet{molina2023model} analyzed the migration choices of individuals and suggested that weather changes, especially those affecting agriculture, directly influence whether people choose to migrate. Another line of research explores social media data to understand public opinions and attitudes towards immigration \cite{potzschke2017migrant,mazzoli2020migrant}. For instance, recent work by \citet{kelling2023analysing} used machine learning algorithms to characterize comments on Facebook in response to refugee-related news articles. Analysis shows that in regions with less hospitable refugee policies, the dominant discourse was more likely to use identity frames and uncivil or misinformation-influenced themes while focusing less on substantive or economic-related frames. In another work, \citet{czymara2023catalyst} analyzed a large dataset of YouTube comments and found that terror attacks increase interest in immigration-related topics and lead to a disproportionate rise in anti-immigrant hate speech, significantly driven by an influx of hateful users rather than a change in sentiment among regular users. Some of the other examples we found include policy analysis \cite{beduschi2017big}, assessing the impact of COVID-19 on global immigration\cite{mcauliffe2021digitalization}, anti-immigrant hate speech analysis \cite{basile2019semeval}, and analysis of stereotypical portrayals of immigrants in online images \cite{olier2022stereotypes}, among other applications. The knowledge generated by studies in this line of research can inform national and international policies.

We also found numerous works with implications in the immigration sector at the national level of governance. For example, \citet{pettrachin2023did} mined spatial and voting data at the 2019 European elections in Berlin. The study suggests voting patterns in various areas were significantly influenced by socio-cultural and historical factors, as opposed to the number of asylum seeker residents. In another example of mining large datasets at the national level, \citet{COCKX2023102306} analyzed a large set of administrative data and found that training programs tailored for unemployed immigrants are effective in improving the overall labour market outcomes in Belgium, suggesting the need for tailored approaches in labour policies. Other examples include studying mental health among newcomers \cite{park2023machine}, asylum claim analysis \cite{barale2023asylex}, anti-immigrant hate speech analysis \cite{bajt2016anti, calderon2020topic}, and immigrant-related political speech analysis \cite{nonnecke2022harass}. Overall, this category of studies enhances national policies, leading to more informed selection processes and improved resource allocation procedures.

We found very few works that use data analytics to inform settlement processes. For example, \citet{khatua2021struggle} and \citet{khatua2022rites} analyzed social media conversations to understand the struggles of settlement. \citet{chen2022influence} used AI methods to assess the impact of education policies on immigrant children's education, and  \citet{voyer2022strange} analyzed the historical corpus of American etiquette books to highlight the experience of social integration. While minimal literature exists, available resources and data analytic techniques remain under-exploited in informing settlement processes.

\vspace{5pt}
\noindent \textbf{Application-Specific Customized AI Tools}: AI research has focused extensively on creating customized solutions for immigration-related challenges, refining these applications with targeted training data to deliver precise outcomes. Note that the utility of the customized AI tools is to ``automate a task'' to improve decision-making rather than ``understand patterns,'' as discussed previously. These systems are often designed to be integrated into broader workflows or processes to enhance efficiency and improve overall results.

AI researchers have extensively investigated the potential of AI tools to enhance international immigration governance. To overcome the adverse effects of language barriers in immigration, a large body of work has focused on building machine translation tools in migration-related contexts, specifically for low-resource languages \cite{cadwell2019more}, although a lot more work on this task is required \cite{macias2020study}. Moreover, AI has frequently been used to forecast migration trends \cite{nalbandian2022advanced}. Several studies have developed such AI technologies within humanitarian programs and displacement response initiatives \cite{beduschi2017big, bither2021automating}. Others have developed models to allocate resources and respond to emergencies related to forced displacement, including predicting the likelihood of individuals overstaying visas in the U.S. \cite{azizi2020artificial}, anticipating future arrivals of refugees and internally displaced persons \cite{doi:10.1080/1369183X.2022.2100546}, predicting optimal settlement locations for migrants in their host countries \cite{bansak2018improving}, forecasting integration trends \cite{juric2022forecasting}, preventing migrant smuggling \cite{nor2022potential}, and legal assistance for refugees \cite{rebolledo2022legal}. Some of this work transcends research, as exemplified by the Foresight Software, developed by the Danish Refugee Council and IBM Research, which has been deployed to predict forced displacement for humanitarian planning \cite{bither2021automating}. 

Our investigation shows that the largest category of AI research applied to immigration-related challenges focuses on building tailored systems for the national level of governance. Some examples of these applications include machine translation \cite{liebling2020unmet}, prediction of migration levels \cite{runfola2022deep}, forecasting policy outcomes \cite{aydemir2022predicting}, legal text processing \cite{rehaag2023luck,nejadgholi2017semi}, language assessment \cite{ramesh2022automated}, public communication for migration policies \cite{sivaperumal2024natural}, social media summarization for inclusive policy-making \cite{kazemi2023ineire}, border security management \cite{khan2021use}, and automated asylum seeker interview \cite{mcnamara2023well}. A full list of applications we found is presented in Table 1.

Similar to data analytics, there is limited research that develops AI tools for applications that improve settlement processes. Existing applications include systems for employment support \cite{uribe2022skill,nikiforos2022machine}, public service accessibility \cite{mariani2022improving}, integration support \cite{lelis2020nadine}, health literacy estimation \cite{tuncer2024estimation}, personal assistants for migrants \cite{wanner2021towards}, assistive tools for overcoming language barriers \cite{nteliou2021digital}, and employment-related skill training \cite{hussain2024conversational}. Yet again, the settlement space would greatly benefit from further AI research.

\vspace{5pt}
\noindent \textbf{General-Purpose AI tools:} This category captures the ad-hoc use of general-purpose AI systems, such as ChatGPT and Google Translate, which are commonly used to enhance efficiency in tasks such as information collection, translation, content generation, and interpretation. Despite the prevalent usage of these tools by the public and in workplaces \cite{butler2023microsoft}, we did not find any study that investigates the use of general-purpose AI tools among newcomers. Therefore, Table \ref{tab:ai_literature} does not include this category. However, we found other evidence of the use of these tools among newcomers. \citet{nedelcu2022precarious} highlighted how migrants heavily rely on publically available digital technologies, specifically information and communication technologies such as mobile phones and search engines, for critical communication and information access. Additionally, in the absence of reliable and customized online information delivery platforms, the ad hoc use of generative AI, such as ChatGPT, is becoming common practice among newcomers and service providers. The Office of the United Nations High Commissioner for Human Rights' latest report on Digital Border Governance notes the growing use of generative AI in the migration sector internationally \cite{DigitalB11:online}. In Canada, an article from CIC News \cite{Howyouca99:online} shared a ChatGPT guide for Canadian newcomers, which has reached over 30,500 shares in a few months. The guide's content, the article's context, and the total number of shares are early signs of an accrued interest in generative AI by newcomers. There is, however, an evident lack of data on this usage among newcomers and SPO staff workers that must be further investigated. 

\vspace{5pt}

\noindent \textbf{Gap in AI Tools for Settlement}: Our examination of the landscape of AI research for immigration demonstrates that most of the developed insights and tools are designed for governance and control at the national and international governance levels. While streamlining administrative efficiency and reinforcing security measures are crucial for maintaining legality and order in immigration procedures, they often prioritize the interest of the state and powerful institutions, potentially at the expense of the individual needs of immigrants. For instance, tools developed for automated border security and asylum claim processing are aimed at optimizing procedural outcomes for national bodies, often with a primary emphasis on risk management and prevention of illegal entry \cite{ajana2015augmented}. In such instances, immigration authorities become intertwined with national defence authorities. While these goals are important, they do not directly address the complex challenges faced by the most vulnerable stakeholders — immigrants themselves — whose lives are impacted by these systems. 

In contrast, areas where AI could potentially empower immigrants for smoother economic, cultural, and social integration, such as personalized education and information delivery tools, remain significantly under-explored. Specifically, early settlement is a crucial, information-intensive phase of long-term integration that necessitates efficient data management and support systems. While customized AI tools could improve personal experiences with immigration, they also promote broader social benefits resulting from improved settlement processes. In the absence of AI tools specifically designed for newcomers, there are substantial risks that they may resort to using general-purpose AI technologies in ad hoc and potentially unsafe ways. This makeshift approach could lead to unreliable or inappropriate solutions, complicating rather than facilitating the integration process for immigrants. 

\vspace{5pt}
\noindent \textbf{An Imperative Shift of Focus for AI Research}: There is a significant need for AI researchers to deeply consider the impact of their work on the distribution of power within immigration governance. The current landscape of research predominantly focuses on enabling international and national stakeholders by providing insights and technical tools. Although this approach can indirectly benefit immigrants, AI research can offer more direct ways of empowering settling newcomers by giving them increased autonomy, specifically for those who have been marginalized. A novel, human-centered approach requires AI applications specifically tailored for use by SPOs and newcomers themselves.

An example of this focus shift can occur in tackling xenophobic rhetoric. While AI research has paid considerable attention to this issue (as evidenced in Table 1), its objective has mostly been limited to anti-immigrant hate speech detection. These tools are necessary for understanding societal attitudes towards immigrants; however, they often focus on passive monitoring and do not translate into direct empowerment or support for the affected individuals. To genuinely support immigrants in the face of such challenges, AI tools should be developed to not only detect xenophobic language but also equip immigrants with resources to counteract and navigate these negative encounters \cite{kiritchenko2021confronting}. Such tools could provide real-time support and advice, legal assistance options, or even educational content to help immigrants respond to and cope with hateful incidents \cite{fraser-etal-2023-makes}. This shift in focus from merely monitoring to actively empowering immigrants represents a critical evolution in AI research, aligning technology more closely with human-centered objectives. To foster this shift of focus, we identify the AI tasks that can be integrated into the settlement sector and aim to ``support and empower settling newcomers.''

\begin{table*}[h!]
\centering
\captionsetup{justification=centering} 

\small 
\renewcommand{\arraystretch}{1.5} 
\begin{tabular}{>{\raggedright\arraybackslash}p{2cm} 
                >{\raggedright\arraybackslash}p{3cm} 
                >{\raggedright\arraybackslash}p{2cm} 
                >{\raggedright\arraybackslash}p{2cm} 
                >{\raggedright\arraybackslash}p{2cm} 
                >{\raggedright\arraybackslash}p{4cm}}
\toprule
\toprule
\textbf{Services Type } & \textbf{Example Application} & \textbf{Output Type} & \textbf{Deployed by} & \textbf{End User} & \textbf{Implications} \\
\midrule
\midrule
Indirect Settlement Services & Decision Support, Resource Allocation & Trends and Insights & National or International Co-ordination Bodies& Government agencies and SPOs & 
\begin{minipage}[t]{\linewidth}
    \begin{itemize}[leftmargin=*]
        \item Provides strategic insights
        \item Supports policy-making
        \item Needs centralized data
    \end{itemize}
\end{minipage} \\
\midrule
Direct Settlement Services & Tailored Information Delivery, Employment Services, Language Training & Predictions, Tailored Information & SPOs & SPO Staff, Newcomers & 
\begin{minipage}[t]{\linewidth}
    \begin{itemize}[leftmargin=*]
        \item Increases efficiency 
        \item Enhances service delivery
        \item Facilitates integration
    \end{itemize}
\end{minipage} \\
\midrule
Decentralized Services & Generic Conversational AI, Generic Machine Translation  & Generic Information & Varied Providers & Newcomers & 
\begin{minipage}[t]{\linewidth}
    \begin{itemize}[leftmargin=*]
        \item Unguarded use
        \item Risk of over-reliance 
        \item Needs AI literacy and critical engagement
    \end{itemize}
\end{minipage} \\
\bottomrule
\bottomrule
\end{tabular}
\caption{Potential Landscape of AI Applications in the Settlement Sector}
\label{tab:table_2}
\end{table*}

\section{A Mapping of AI Tasks for Settlement }

In this section, we first explain the Canadian settlement journey and then suggest AI applications that can facilitate information-processing tasks in this heavily burdened sector.

\subsection{The Canadian Newcomer Settlement Journey}

\vspace{5pt}
The typical newcomer’s journey to Canada has two phases, each requiring different settlement services for their changing needs: the pre-arrival and post-arrival phases. The post-arrival phase is further divided into two periods with distinct needs: the short-term settlement needs and the long-term settlement needs. Importantly, settlement services are only offered to eligible newcomers (permanent residents, refugees, and temporary residents from specific streams) \cite{Immigrat27:online}
; others may seek the same type of services at a cost, at charitable organizations, or on their own time. While the settlement journey remains very similar across Canada, the province of Québec offers its own settlement services under a special Canada-Québec accord \cite{CanadaQuebec1991}.

The pre-arrival phase begins once newcomers receive confirmation of their immigration status. This phase generally takes place zero to three months prior to their arrival in Canada. At this stage, newcomers' most pressing needs are to receive information and orientation services, such as guidance regarding Canada's climate and the general lifestyle to prepare for.  Moreover, they can receive employment-related information such as foreign credential recognition or referrals to employment opportunities. They can also be preemptively connected to SPOs to continue their settlement and integration post-arrival. Pre-arrival settlement services can be offered in-person in select countries (China, India, Morocco, and the Philippines) and remotely through video conference \cite{CanadaImmigration}.

Short-term post-arrival services focus on newcomers' pressing needs in their first three months in Canada, such as clothing, food, well-being and mental health, security, housing, employment, digital access, translation support, and assistance to fill important documents. Settlement workers at SPOs assess these needs through \textit{Needs and Assets Assessment and Referrals Services} (NAARS) or through an informal discussion if the newcomer requests help for a single settlement service \cite{Outcomes2021}.

Long-term post-arrival services focus on newcomers' successful integration after their first three months in the country. Newcomers are encouraged to build a network through community connections and to integrate in Canada through formal language classes or informal training through conversation circles. In some cases, newcomers may require further orientation information or employment-related services to adapt to Canadian life or find a job aligned with their prior expertise. At this stage, newcomer families will also focus on finding schools for their children \cite{Outcomes2021}.

The timeline and journey described above apply to newcomers in general. For resettled newcomers such as refugees, the timeline might vastly differ as they may require more assistance. For instance, refugees require an extensive NAARS process to assess their complex needs. Furthermore, refugees may, in some cases, deal with severe trauma. As such, their short-term post-arrival needs may be extended with urgent care for health and well-being. Thus, their integration and long-term settlement needs may be delayed compared to the average newcomer. However, resettled newcomers will go through the same process as the average newcomer once their urgent needs are met \cite{Outcomes2021}.

\begin{table*}[h!]
\centering
\small 
\begin{tabular}{>{\raggedright\arraybackslash}m{3.5cm} >{\raggedright\arraybackslash}m{6cm} >{\raggedright\arraybackslash}m{6cm}}
\toprule
\textbf{Service (Fund \%)} & \textbf{Example AI Application } & \textbf{Core AI Tasks} \\
\midrule
Language Training and Assessment (30) & 
\begin{itemize}
    \item Assisting language training via voice-enabled and text-processing AI tools
    \item Language proficiency assessment through automated testing and scoring systems
    \item Automatic creation of personalized study plans
    \item Automatic learning resource recommendations
\end{itemize}
& 
\begin{itemize}
    \item Machine translation
    \item Dialogue systems
    \item Automatic feedback and corrections
    \item Assessment algorithms
    \item Recommendation systems
\end{itemize}
\\ \hline
Information and Orientation Services (19) & 
\begin{itemize}
    \item Automated, personalized information delivery in key areas
    \item AI-driven guides on finding further information and digital skill training
\end{itemize}
& 
\begin{itemize}
    \item Information retrieval
    \item Question answering systems
    \item Conversational AI
    \item Personalized recommendation systems
\end{itemize}
\\ \hline
Needs and Assets Assessment and Referral Services (NAARS) (11) & 
\begin{itemize}
    \item Automating routine data processing
    \item Analyzing client information to predict needs
    \item Automatic referral generation based on the needs and assets 
\end{itemize}
& 
\begin{itemize}
    \item Data processing automation
    \item Predictive algorithms
    \item Matching algorithms
    \item Recommendation systems
\end{itemize}
\\ \hline
Support Services (9) & 
\begin{itemize}
    \item AI-coordinated childcare and transportation scheduling
    \item Adaptive technologies for disability support
    \item Automated translation and interpretation systems
    \item Expert-in-the-loop mental health support
    \item Fraud and scam detection 
\end{itemize}
& 
\begin{itemize}
    \item Scheduling algorithms
    \item Machine translation
    \item Conversational agents
    \item Recommendation systems
    \item Classification
\end{itemize}
\\ \hline
Employment-Related Services (9) & 
\begin{itemize}
    \item Automated job matching systems
    \item Resume and cover letter assistance tools
    \item Online training modules for skill development
    \item Personalized career advice and planning
\end{itemize}
& 
\begin{itemize}
    \item Matching algorithms
    \item Generative AI
    \item Conversational assistants
    \item Personalized recommendation systems
\end{itemize}
\\

\bottomrule
\end{tabular}
\caption{Potential AI applications in direct settlement services based on service types outlined in \cite{part1outcome:online}.}
\label{tab:table_3}
\end{table*}

\subsection{Potential AI Solutions for Settlement }

Table \ref{tab:table_2} categorizes potential AI applications in settlement services into three distinct, high-level categories, depending on who would deploy and use the AI tool: 1) \textit{AI for indirect settlement services} deployed by international and national co-ordination bodies to support SPOs, 2) \textit{AI for direct settlement services} deployed by SPOs to support newcomers, and 3) \textit{Decentralized AI tools} deployed by varied providers and used by newcomers. The third category is commonly used by temporary residents who are either not eligible for direct services or eligible clients who have no access to services due to limited resources. 

The first category characterizes AI applications deployed by international or national stakeholders to identify trends and generate insights for use by government agencies and SPOs. This includes decision support and resource allocation systems. For example, such applications might analyze historical data collected during settlement to identify the success factors or barriers of social and economic integration. These applications require significant investments in coordinated research and development, including centralized data infrastructures. Moreover, these tools have strategic implications as they inform policy-making and provide a big-picture view that can aid in resource allocation. In the second category, AI-powered tools are deployed by the SPOs to provide tailored information delivery, assist in employment services, and facilitate language training to directly serve newcomers. These applications increase efficiency and enhance experiences with service delivery. Along the third category, newcomers who do not have access to settlement services might use generic conversational AI and machine translation services deployed by various providers to collect information. Although these tools are cost-effective and accessible, the high risks of unsafe use and over-reliance on technology must be considered \cite{Building33:online}, especially in the use of general-purpose conversational AI tools \cite{weidinger2021ethical,amershi2019guidelines}.

Further, Table \ref{tab:table_3} maps the subcategories of direct services to AI applications and the corresponding core AI tasks. The full list of subcategories is outlined in \cite{part1outcome:online}. From this list, we excluded the \textit{Resettlement Assistance Program}, which focuses on the immediate post-arrival needs of resettled refugees and other eligible clients, and the \textit{Community Connection} services, which help clients learn about and integrate into their local communities. In our engagement with SPOs, we learned that expert insight and personal connection are extremely important for both of these services, making them far from ideal cases for automation. The following section elaborates on use cases where AI can be reliably and responsibly integrated into SPOs' existing workflows.  

\vspace{5pt}

\noindent \textbf{Language Assessment and Training }: With 30\% of the total settlement funding, these programs provide clients with English and French language training. Such services can potentially incorporate Intelligent Computer Assisted Language Learning (ICALL) tools and techniques, which have largely utilized machine learning and Natural Language Processing (NLP) techniques to enhance second language learning. These tools are used to identify errors, provide feedback, and assess language skills \cite{22systematic}. While the integration of AI in language training promises a personalized, flexible and inclusive educational environment, fears exist that AI might dominate student learning and dictate learning processes without consent \cite{pokrivcakova2019preparing}. In designing such tools, teachers and learners should be in control of AI tools so that they fully understand and organize educational activities.

\vspace{5pt}
\noindent \textbf{Information and Orientation}: Despite accounting for the second largest amount, 19\%, of settlement funding, these services are the most accessed in terms of unique clients \cite{part1outcome:online}. Their aim is to provide newcomers with essential information regarding housing, healthcare, education, finances, law and justice, crucial documents, digital skill training, and ways to access further information. Integrating core NLP tasks such as information retrieval, question-answering agents, conversational agents, and recommendation systems can enhance the effectiveness of these services and streamline information provision. The integration of such tools in SPOs' existing workflows might be challenging since it requires staff training and constant maintenance of developed technologies to ensure that they reflect the most recent and appropriate information. 

\vspace{5pt}

\noindent \textbf{Needs and Asset Assessment and Referral Services (NAARS)}: 
Receiving 11\% of settlement funding, assessment services are crucial for determining clients' specific assets and needs and providing targeted referrals to appropriate supports. While the complexity of these services necessitates expert judgment, there are promising opportunities to augment NAARS services through automating routine data processing tasks, analyzing client information to predict needs, and generating referral options. Additionally, machine learning algorithms could be used to continuously improve service recommendations by learning from outcomes and feedback. This integration of AI would not replace the essential human element but would instead free up valuable time for staff to focus on more complex and sensitive aspects of client interaction, thereby streamlining the process and enhancing service delivery.

\vspace{5pt}

\noindent \textbf{Support Services}: With 9\% of the settlement funding, these services address the day-to-day challenges faced by newcomers, including childcare, transportation, disability services, translation and interpretation, digital solutions, and short-term counselling. AI-coordinated scheduling systems can efficiently manage appointments and logistics for childcare and transportation, reducing administrative burdens and allowing staff to focus more on client interaction and personalized service. Adaptive technologies for persons with disabilities may include AI-driven avatars for ASL interpretation and screen readers that improve accessibility to digital content. Additionally, automated translation and verbal interpretation systems help bridge language barriers, especially in critical settings such as medical appointments or legal consultations. AI-assisted tools for short-term counselling might also provide support to newcomers, although they require human oversight to ensure sensitivity and appropriateness of responses. In this setting, SPO staff oversee AI applications to ensure they meet clients' diverse and specific needs, embodying an expert-in-the-loop approach that maintains essential human control in service delivery.

\vspace{5pt}

\noindent \textbf{Employment-Related Services}: Receiving 9\% of the total settlement funding, these services prepare newcomers for the labour market. Automated job matching systems can optimize job search and training processes in alignment with clients' skills and qualifications. Writing assistant tools can help clients present their qualifications effectively in resumes and cover letters tailored to specific job markets and employer expectations. Moreover, online training modules powered by AI can provide personalized skill development pathways \cite{yang2024social}, which are crucial for newcomers adapting to new job environments. Human oversight by SPO staff is necessary to tailor AI recommendations to individual circumstances, ensuring that advice and opportunities are culturally and contextually appropriate. This expert-in-the-loop approach ensures that the human element remains at the core of service provision while AI handles the data-intensive tasks. Expert oversight is especially crucial in this context because AI systems can be biased and may make stereotypical recommendations in employment \cite{FAccT,nadeem-etal-2021-stereoset}, necessitating vigilant review to ensure fairness and accuracy in career guidance.

\subsection{Data in Settlement Sector }
Data, specifically data collection practices, has an increasingly important role in every aspect of migration. Data collection can occur pre- and post-arrival. Before settlement, data from publicly available resources, such as Social Networking Sites (SNSs), have been used to map and forecast migration pathways during migration crises \cite{mazzoli2020migrant}. State authorities are known to collect general data on migrants for immigration management purposes, yet it often translates into law enforcement usage, specifically when the datasets are shared between immigration authorities and national defense or law enforcement authorities \cite{vavoula2020consultation}. This practice is known to be a violation of the `purpose limitation principle,' which dictates that data collection must only be for a ''[...] specified, explicit and legitimate purpose, and must not be further processed in a manner that is incompatible with that purpose.'' \cite{vavoula2020consultation, montaldo2020fight}. 

Post-arrival data is collected at a smaller scale. For instance, local SPOs collect data on newcomers’ needs and assets to refer them to specialized organizations \cite{ISC}. This process is sometimes centralized, as is the case with Immigrant Services Calgary. Some of the same data is also shared with the IRCC to produce annual reports on settlement needs \cite{immigration_settlement_2022}. While data collection is currently limited in the post-arrival settlement journey, it is imperative to apply the ``purpose limitation principle'' for any future data manipulation, especially for AI applications, to avoid any dataset transfer between immigration authorities and law enforcement agencies.

Some migrants, on the other hand, use digital solutions and data from unofficial channels to facilitate their settlement journey \cite{galis2022analog}. Smartphones become an important tool for irregular migrants to share their geolocation with families back home and to obtain data on the preferred pathways shared in large group chats on social networks. The data on these unofficial channels quickly become vital to a successful settlement journey for irregular migrants. Such migrants also highly prioritize data anonymity to avoid State surveillance on their use of irregular routes \cite{sanchez2018connected}.  

The challenges presented by the limited, sensitive, and decentralized nature of data in the settlement sector pose barriers to the development of AI solutions. Access to such data is restricted, complicating efforts to employ AI effectively in this context. To overcome these hurdles, it is crucial to explore innovative approaches, such as the use of synthetic data or alternative data sources that can enrich AI applications within the settlement sector \cite{jordon2022synthetic}. Additionally, enhancing mechanisms for data sharing while ensuring robust data anonymization and privacy-friendly AI practices is essential \cite{li2020review}. 

\subsection{Responsible AI Development}
Developing AI tools for the immigrant settlement sector necessitates rigorous adherence to the principles of responsible AI due to its direct impact on the highly sensitive and personal aspects of individuals' lives. Here, we introduce the available guidelines and highlight the aspects of responsible AI that are specifically important for this sector.  

\vspace{5pt}
\noindent \textbf{Principled AI Design}:  The \textit{Montreal Declaration for a Responsible Development of Artificial Intelligence} \cite{Montreal-Dec} and the \textit{Directive on Automated Decision-Making} \cite{Directive} in Canada provide frameworks to guide AI development. The \textit{Montreal Declaration} emphasizes the importance of well-being, autonomy, and justice and supports societal advancements while prioritizing ethical considerations. It also emphasizes democratic participation in AI development, ensuring diverse community involvement, which is particularly relevant in the culturally-rich context of immigrant services. Similarly, the Canadian \textit{Directive on Automated Decision-Making} imposes standards for the use of AI in government operations, mandating impact assessments, transparency, and accountability. These directives require AI systems to be tested for accuracy, fairness, and potential biases, with a clear process for human intervention and the rectification of errors.

\vspace{5pt}
\noindent \textbf{Preventing Information Hazards}: Given the sensitive nature of settlement data — ranging from legal status to health information — data privacy is a cornerstone of designing AI tools for the sector to prevent the misuse or exposure of individuals' personal information. Data must be collected with explicit consent, ensuring that individuals are fully informed about how their information will be used and have control over its use. Data minimization should also be practiced to ensure that only necessary information is collected to reduce the risk of harm. Moreover, robust data protection measures must be implemented to secure personal data against breaches or unauthorized access. These measures are particularly crucial in the immigrant settlement sector, where the stakes of data misuse are high, potentially affecting individuals' legal statuses, access to services, and overall safety.

\vspace{5pt}
\noindent \textbf{Fairness and Non-Discrimination}: These principles are specifically relevant when developing AI tools for the settlement sector, which serves diverse populations, often including vulnerable groups such as refugees and individuals facing significant cultural, linguistic, and socio-economic adjustments. Care should be taken so that deployed AI systems do not perpetuate biases or inequities ingrained in their training sets. Additionally, it is critical to integrate intersectional considerations into designing these systems. Neglecting to incorporate an intersectional approach can lead AI systems to overlook the complex realities of individuals who face multiple forms of discrimination simultaneously. This oversight can exacerbate the unique challenges and barriers faced by those at the intersections of multiple marginalized groups within the newcomer community, leading to adverse impacts on their ability to integrate successfully and equitably.

\vspace{5pt}
\noindent \textbf{Trustworthy AI Deployment}: Transparency and accountability are other principles that become crucial in building trust among users and affected communities to ensure that they understand the tools that influence their integration process, such as language training, job placement, and legal assistance. SPOs are widely recognized as reliable resources for newcomers, reinforcing their reputation as vetted and credible sources. As these organizations integrate AI into their services, they should invest in processes that guarantee human supervision of the technology. In cases where newcomers directly interact with the system deployed by settlement services, the established credibility of SPOs may enhance the perceived reliability of their AI systems, leading to over-reliance on AI outputs. To mitigate such risks, settlement services need to invest in transparent communication about the limitations of such systems and their flaws, including potential misinformation.

\section{Call to Action for Multidisciplinary Research}

In this paper, we emphasize the potential for AI to transform the settlement phase of the immigration journey and the significant opportunities for AI researchers to make a meaningful impact. The following call to action, guided by the OECD's AI Principles \cite{OECD-AI-Principle}, outlines our recommendations for the essential steps that researchers, policymakers, and settlement organizations should take to address these challenges:

\subsection{Role of AI Research}

\vspace{5pt}
\noindent \textbf{Fostering a Research Community:} AI researchers should recognize the settlement sector as a crucial area of application and dedicate resources to support this sector. This requires defining clear tasks, organizing technical workshops, and building research communities similar to those established in legal, financial, and medical domains.

\vspace{5pt}
\noindent \textbf{Designing AI to Empower Newcomers:} We advocate for a shift in research priorities such that AI tools are designed to empower newcomers rather than reinforce systemic power dynamics. 

\vspace{5pt}
\noindent \textbf{Engaging with Communities of End Users}: AI researchers should closely work with SPOs and newcomers to understand this sector's assets and needs in the early stages of designing AI-enabled solutions. AI should aim to augment the outreach and impact of existing support structures and settlement service providers rather than replace them.
    
\vspace{5pt}
\noindent \textbf{Collaborating with Social Scientists:} Social scientists have long studied settlement and successful integration. Future AI research should draw on the specialized expertise of political scientists, sociologists, and psychologists to develop comprehensive and culturally sensitive solutions.

\subsection{Role of Service Provider Organizations}

\vspace{5pt}
\noindent \textbf{Leveraging AI for Improved Efficiency:} SPOs should actively embrace new technologies for increased efficiency and outreach. Through collaboration with AI researchers, SPOs can prototype, test, and deploy AI solutions for language training, skill development, and employment matching. In doing so, SPOs can create a more effective support structure that accelerates newcomers' economic and social integration, ultimately benefiting society as a whole.

\vspace{5pt}
\noindent \textbf{Investing in Data Infrastructure:} Representative, anonymized, and ethically sourced datasets on newcomers' needs, challenges, and successes are at the core of building responsible AI solutions tailored to the settlement sector. SPOs should invest in secure and standardized data collection, storage, and sharing systems to enable collaboration with AI researchers and developers. 

\vspace{5pt}
\noindent \textbf{Investing in AI Literacy for Newcomers and Staff:} SPOs should prioritize investments in AI literacy programs for both newcomers and their staff. For newcomers, these programs should include accessible workshops and resources that demystify general-purpose AI technologies, highlighting their potential and limitations. Additionally, creating a culture of AI awareness within organizations with specialized training for staff is essential to facilitate the adoption of new technologies. 

\vspace{5pt}
\noindent \textbf{Preserving the Human Touch:} SPOs should strive to maintain a balance where technology enhances service delivery without diminishing the personal touch that is fundamental to newcomers' successful support and integration. The empathy, understanding, and personal connections that staff provide cannot be fully replicated by technology, and AI should be seen as a tool to augment, not replace, the invaluable human element of their services.


\subsection{Role of Government} 

\vspace{5pt}
\noindent \textbf{Coordination of Research and Innovation:} Governments play a crucial role in promoting innovative AI ecosystems and coordinating activities among various actors. Long-term public investment in representative datasets that respect privacy can drive innovation while addressing technical, social, legal, and ethical challenges. An agile policy environment supported by data access, computational infrastructure, and stakeholder collaboration can facilitate responsible AI use and ensure a fair labour market transition. International cooperation is also vital to establishing global standards and partnerships.

\vspace{5pt}
\noindent \textbf{AI Regulation in the Settlement Sector:} Government regulation may need to restrict AI applications in areas with profound legal and personal implications. The implementation of laws to govern AI in these contexts could guarantee that only accurate information is delivered, thus preventing potential misinformation or other harmful consequences.


\section{Conclusion and Future Work}

Through participatory research and scanning the AI research literature, we laid out a vision for AI-empowered solutions to alleviate the challenges faced by the settlement sector. We urge AI researchers to build upon our work and collaborate across disciplines to advance this vision for developing human-centred, settlement-focused AI tools that underscore empowerment, inclusivity, and safe use of AI while expanding efficient service provision. Although this paper focuses on Canada, our findings and recommendations largely apply to other contexts and host countries. 

While we have discussed ample caution on data privacy and fairness issues, a broader perspective informed by human-AI interaction and sociotechnical literature could further enrich the discussion on potential pitfalls. This includes how AI tools may be used or resisted by different stakeholders, the potential impact on service quality and human agency, and the shifts in power dynamics within the system. Additionally, future work should assess the risks of subjecting immigrants to heavy data surveillance compared to the native population through the integration of AI tools. Addressing these aspects can help ensure that AI solutions are implemented with a keen awareness of the social contexts and power dynamics among immigrants, service providers, and other stakeholders.

\nocite{*} 
\bibliography{custom}

\end{document}